\documentclass[10pt,amsmath,amssymb,nofootinbib,twoside,twocolumn,superscriptaddress,floats,floatfix,aps,prd,preprintnumbers]{revtex4-2}

\usepackage{graphicx,epsfig}
\usepackage{amsthm,amsfonts}
\usepackage{mathrsfs}
\usepackage{bm}
\usepackage[inline]{enumitem}
\usepackage{tensor}
\usepackage[linktocpage]{hyperref}
\usepackage[capitalise]{cleveref}
\usepackage[dvipsnames]{xcolor}
\usepackage{url}
\usepackage{xspace}
\usepackage{comment}
\usepackage{ dsfont }
\usepackage[normalem]{ulem}
\usepackage{subfigure}
\hypersetup{
    colorlinks=true,
    linkcolor=red,
    citecolor=blue,
    filecolor=black,
    urlcolor=blue,
    pdfauthor={M. Della Rocca, P. Pani, and A. P. Sanna},
    pdftitle={}
}

\newcommand{\dd}{\text{d}}                              % Differential symbol

\newcommand{\ii}{\text{i}}
\newcommand{\ee}{\text{e}}

\renewcommand{\Re}{\mathrm{Re}}
\renewcommand{\Im}{\mathrm{Im}}
\definecolor{mygreen}{rgb}{0., 0.60, 0.}

\newcommand{\sapienza}{Dipartimento di Fisica, Sapienza Università 
	di Roma, Piazzale Aldo Moro 5, 00185, Roma, Italy}
\newcommand{\infn}{INFN, Sezione di Roma, Piazzale Aldo Moro 2, 00185, Roma, Italy}

\allowdisplaybreaks

\begin{document}

\author{Matteo~Della~Rocca}
\email{matteo.dellarocca@uniroma1.it}
\affiliation{\sapienza}\affiliation{\infn}

\author{Paolo~Pani}
\email{paolo.pani@uniroma1.it}
\affiliation{\sapienza}\affiliation{\infn}

\author{Andrea~P.~Sanna}
\email{asanna@roma1.infn.it}
\affiliation{\infn}

\title{Can One Hear the Absence of a Singularity?
}

\begin{abstract}
 One of the central expectations of quantum gravity is the resolution of the curvature singularities predicted by General Relativity, potentially giving rise to effective black-hole geometries that are regular inside the horizon. 
 We show that, for a very broad class of regular black holes including several known solutions, the quasinormal-mode spectrum in the large-overtone limit differs discontinuously from that of a Schwarzschild black hole, even when the regularization parameter is arbitrarily small. The real part of the quasinormal frequencies vanishes asymptotically, rather than approaching a nonzero constant, as generically occurs in General Relativity and other theories predicting black holes with curvature singularities. The asymptotic black-hole spectrum thus retains a distinctive imprint of singularity resolution.
 We establish these results analytically through a systematic formulation of the monodromy method and confirm them numerically. We interpret this phenomenon as a manifestation of black-hole spectral instability, whereby arbitrarily small deviations from vacuum General Relativity produce nonperturbative changes in the large-overtone spectrum.
\end{abstract}

\maketitle

%%%%%%%%%%%%%%%%%%%%%%%%%%
{\bf\em Introduction.}
% \label{sec:Intro}
%%%%%%%%%%%%%%%%%%%%%%%%%%
As established by famous theorems due to Penrose and Hawking~\cite{Penrose:1964wq,Hawking:1970zqf,Senovilla:2014gza}, curvature singularities are generically predicted by General Relativity~(GR) and mark one of the sharpest limitations of the theory. Their resolution is widely expected to require quantum gravity, although the mechanism by which this occurs remains unknown.
A broad class of phenomenological models implements this expectation through effective geometries in which the Schwarzschild singularity is replaced by a regular core (see Ref.~\cite{Carballo-Rubio:2025fnc} for a recent review), while deviations from the GR solution outside the horizon can be arbitrarily small, naturally suppressed by $\ell_{\rm P}/M$, where $\ell_{\rm P}$ is the Planck length and $M$ is the gravitational radius, in $G=c=1$ units henceforth adopted. This raises a fundamental
question: \emph{can the absence of a singularity leave an imprint on the spectrum of a black hole (BH), even when its exterior geometry is arbitrarily close to Schwarzschild?}

Quasinormal modes (QNMs)~\cite{Vishveshwara:1970zz,Press:1971wr,Teukolsky:1973ha,Chandrasekhar:1975zza,Kokkotas:1999bd,Berti:2009kk} provide a particularly intriguing arena in which
to address this question. Besides governing the linear relaxation of
perturbed BHs, their highly damped spectrum has long been suspected to encode information of a more fundamental nature. For a Schwarzschild BH, the asymptotic QNM frequencies display the remarkable  structure~\cite{Motl:2003cd,Andersson:2003fh} 
\begin{equation}
 \omega_n \sim T_{\rm H}\ln 3
 - \ii\,2\pi T_{\rm H}\, n,
 \qquad n\rightarrow\infty ,
 \label{eq:SchwarzschildQNMAsymptotic}
\end{equation}
where $T_{\rm H}$ is the Hawking temperature
and $n$ the overtone number.
The appearance of $T_{\rm H}$ and of the $\ln 3$ factor suggested a link with Bekenstein's proposal of an evenly spaced BH area spectrum~\cite{Bekenstein:1974jk}: via Bohr's correspondence principle, Hod identified ${\rm Re}[\omega_n]$ with the BH transition frequency, obtaining an area quantum proportional to $\ln 3$~\cite{Hod:1998vk}.
Maggiore argued instead that the relevant quantity is the proper frequency $\sqrt{{\rm Re}[\omega_n]^2+{\rm Im}[\omega_n]^2}$, which is controlled by the spacing of ${\rm Im}[\omega_n]$ and yields Bekenstein's $\Delta A=8\pi \ell_{\rm P}^2$~\cite{Maggiore:2007nq} (see also Ref.~\cite{Cadoni:2021jer}).

Highly damped modes are qualitatively different from the low-lying QNMs relevant to conventional ringdown~\cite{Berti:2025hly}. Their analytic structure in the complex radial plane is efficiently captured by the monodromy method~\cite{Motl:2003cd,Andersson:2003fh,Das:2004db,Natario:2004jd}, which has recently received renewed attention as a way of extracting information about BH interiors and their quantum modifications from asymptotic QNM spectra~\cite{Hartnoll:2026vhu,Coviello:2026tol,Coviello:2026hdc}.
This raises an apparent tension: \emph{if a modification of the geometry is confined to an arbitrarily small region around the classical singularity, should its spectral imprint disappear continuously as that region shrinks to zero?}

We show that, in general, it does not: increasing the overtone number magnifies \emph{arbitrarily small} deviations from the Schwarzschild geometry outside the horizon, eventually producing nonperturbative changes in the asymptotic spectrum.
For a very broad class of asymptotically Schwarzschild regular BHs, including several well-known geometries, resolving the central singularity changes the large-overtone QNM spectrum discontinuously:
for any nonzero value of the regularization parameter, \emph{however small}, the asymptotic real part satisfies
\begin{equation}
\Re[\omega_n]\rightarrow 0 ,
 \qquad n\rightarrow\infty , 
 \label{eq:wR}
\end{equation}
rather than approaching the Schwarzschild value
$T_{\rm H}\ln3$. 
Thus, the limits in which the regularization is removed and the overtone number is taken to infinity do not commute.
This also occurs for Reissner--Nordstr\"om~\cite{Motl:2003cd,Andersson:2003fh,Berti:2003zu} and Kerr BHs~\cite{Keshet:2007nv,Keshet:2007be,Daghigh:2010wm}, when taking the zero-charge and zero-angular-momentum limits, respectively.
Remarkably, the vanishing of $\Re[\omega_n]$ in the highly damped limit is not a generic consequence of perturbing the Schwarzschild geometry: theories predicting BHs with curvature singularities generically retain a qualitatively different asymptotic spectrum. This suggests that UV corrections to GR lead to Eq.~\eqref{eq:wR} only if they resolve the central singularity.
We establish these results analytically by developing a systematic monodromy treatment of the highly damped spectrum and confirm them through high-precision numerical calculations. The phenomenon provides a sharp
example of BH spectral instability~\cite{Nollert:1996rf,Barausse:2014tra,Jaramillo:2020tuu,Cheung:2021bol,Berti:2022xfj}: an arbitrarily small modification of the geometry can produce nonperturbative changes in the spectrum at sufficiently large overtone number. In this sense, highly damped QNMs ``hear'' a property invisible to the exterior Schwarzschild limit---the absence of the central singularity.
%

\begin{figure}
    \centering
    \includegraphics[width=\linewidth]{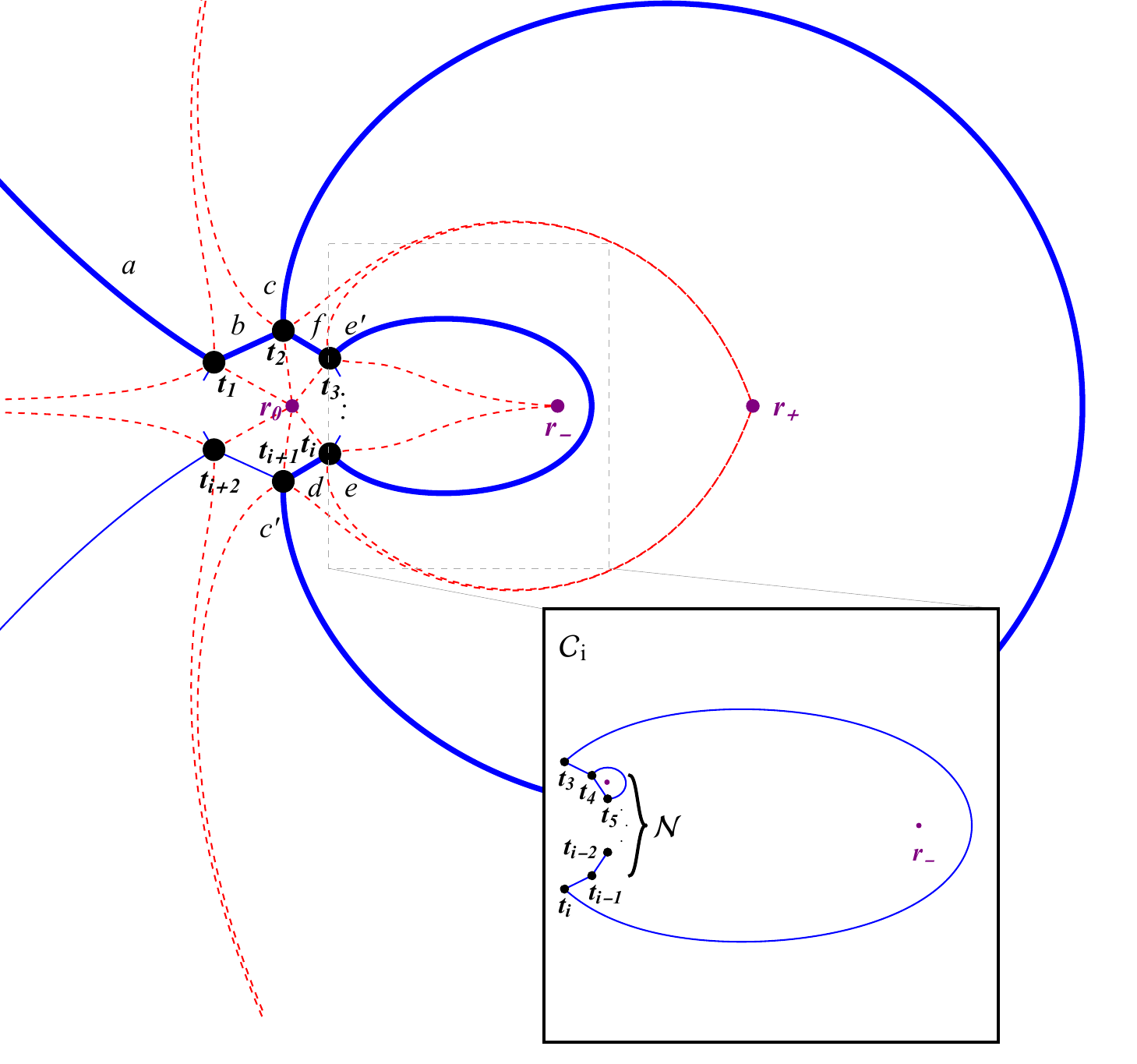}
    \caption{Sketch of the Stokes/anti-Stokes network for a generic, nondegenerate, two-horizon BH. Blue solid (red dashed) lines denote anti-Stokes (Stokes) lines, while black dots mark the turning points of $Q^2$, clustered around a pole $r_0$ of $H$. The outer and inner horizons are denoted by $r_+$ and $r_-$, respectively. The contour $\Lambda$ used to perform the computation is highlighted by thick blue lines. Inset: contour $\mathcal{C}_i$ enclosing the inner horizon and, possibly, additional poles (purple dot). For a nondegenerate, single-horizon BH the contour $\mathcal{C}_i$ is absent.
    }
\label{fig:SketchContour}
\end{figure}

%--------------------------------------------------------------------------------
{\bf\em Highly damped modes combining monodromy and WKB methods.}
%--------------------------------------------------------------------------------
We consider the asymptotic QNM spectrum of a massless scalar field $\Phi$ propagating in a generic static, spherically symmetric, asymptotically flat spacetime, described by the line element
\begin{equation} \dd s^2=-F(r) \dd t^2 + B(r)^{-1} \dd r^2 + r^2 \left(\dd\theta^2+\sin^2\theta\dd \phi^2\right) \, . 
\end{equation}
Expanding the scalar field in spherical harmonics $Y_{\ell m}$ and Fourier modes, $\Phi=r^{-1}\sum_{\ell m} \ee^{-\ii\omega t}\, Y_{\ell m}(\theta,\phi)(FB)^{-1/4}\, \Psi(r)$,
 the radial master function satisfies the Schr\"odinger-like equation
\begin{align} 
\frac{\dd^2 \Psi}{\dd r^2} &+ R(r) \Psi = 0\,,\hspace{0.2cm} \label{eq:EqPerturbationGeneral} \\
\label{eq:R} R=\frac{\omega ^2}{H}-\frac{\ell (\ell +1)}{r^2 B}&+\frac{3 H'^2}{16 H^2}-\frac{\left(r^2H'\right)'}{4r^2H}\,,
\end{align} 
with $H\equiv B F$ and prime denoting differentiation with respect to $r$.
For highly damped modes, i.e., $|\omega| M\gg1$, we solve Eq.~\eqref{eq:EqPerturbationGeneral} within the Wentzel-Kramers-Brillouin (WKB) approximation after analytically continuing the radial coordinate $r$ to the complex plane~\cite{Andersson:2003fh}. The two independent WKB solutions are~\cite{Berry:1972na} 
\begin{equation} 
f_{1, 2}^{(t)}(r) = \frac{\ee^{\pm \ii \gamma_{t}(r)}}{\sqrt{Q(r)}}\, , \qquad \gamma_{t}(r) \equiv \int_t^r Q(\tilde r) \, \dd \tilde r \, ,
\label{eq:WKBgeneralSolution}
\end{equation} 
where $t$ denotes a reference turning point, i.e., a root of $Q^2$. 
Since the behavior near the turning points is crucial to the analysis, the function $Q$ must be chosen appropriately. 
The standard WKB ansatz $Q^2=R$ may not reproduce the correct local behavior of the solution to \cref{eq:EqPerturbationGeneral} near a pole of $H$, around which the turning points accumulate in the high-frequency limit~\cite{Langer:1937qr,Andersson:2003fh}.
If $H$ has a pole of order $2N>0$ at $r=r_0\in\mathbb{C}$, the metric functions generically behave at leading order as
\begin{equation}
B\sim(r-r_0)^{-z}\,,\qquad
F\sim(r-r_0)^{-2N+z}\,,
\label{eq:metric_expansion_r0}
\end{equation}
as $r\to r_0$. 
Note that $r_0=0$ for singular BHs~\cite{footnote}.
As shown in the Supplemental Material~\ref{app:Integral_phase}, for a broad class of metric functions the angular-momentum contribution to Eq.~\eqref{eq:R} is subleading near the pole, and the Langer-type modification $Q^2=R-1/[4(r-r_0)^2]$ correctly reproduces the local behavior of $\Psi$. In particular, this holds for any $z>-2$ when $r_0\neq0$, and for any $z>0$ when $r_0=0$ and $N\neq2$. Special limiting cases, in which additional terms contribute at the same leading order, are discussed in the Supplemental Material. Our framework thus encompasses a large class of singular (i.e., $r_0=0$) and regular (i.e., $r_0 \ne 0$) BH geometries, including the Schwarzschild solution (corresponding to $z=N=1$) and the regular BHs considered below.
As $Q=\sqrt{Q^2}$ is multivalued, branch cuts must be introduced in the complex $r$-plane to define a single-valued branch of $Q$. Following Ref.~\cite{Andersson:2003fh}, we choose the branch such that $Q(r) \sim \omega$ as $r \to \infty$. With this choice, $f_1$ corresponds to a purely outgoing-wave solution at spatial infinity. This identification is made along an anti-Stokes line, where $Q(r)\dd r$ is real by definition, so that WKB solutions are purely oscillatory~\cite{Andersson:2003fh,Berry:1989uni}.   
The QNM boundary condition at infinity, therefore, requires $\Psi_{\rm{QNM}}=f_1$ along an asymptotic anti-Stokes line~\cite{Vishveshwara:1970zz}.
At the horizon, the QNM boundary condition instead requires $\Psi$ to be an eigenfunction of the monodromy matrix associated with a closed contour around the horizon~\cite{Motl:2003cd}. After one entire loop, it therefore transforms as $\Psi_{\rm QNM}= \mathcal{E} f_1$, where $\mathcal{E}$ is the corresponding monodromy eigenvalue. 
As shown in the Supplemental Material~\ref{app:path_existence}, for a wide class of BHs, one can construct a loop $\Lambda$ composed of anti-Stokes lines connected to infinity. A sketch of this contour is shown in \cref{fig:SketchContour}.
Starting from $\Psi=\Psi_{\rm QNM}=f_1$ at infinity (point $a$ in \cref{fig:SketchContour}), we analytically continue it along $\Lambda \equiv a \to b \to c \to c' \to d \to e \to e' \to f \to c \to b \to a$, taking into account the Stokes phenomenon~\cite{Andersson:2003fh,Berry:1989uni}. Upon returning to point $a$, the solution takes the general form $\Psi\to \mathcal{A} f_1+\mathcal{B}f_2$~\cite{Andersson:2003fh}.
Since $\Psi_{\rm QNM}$ must be proportional to $f_1$, the asymptotic QNM spectrum is determined by the condition $\mathcal{B}=0$.
A central result of this work, derived in the Supplemental Material~\ref{subsubsec:DerivationMonodromyGeneralContour}, is that for the broad class of solutions considered here, this condition takes a remarkably simple and universal form
\begin{equation}
\ee^{2\ii \Gamma_e} = 1- \left(1 + \ee^{-2\ii \gamma} \right) \left(1 + \ee^{2 \ii \gamma} \right) \left(1 + \ee^{-2 \ii \Gamma_i} \right)\, .
\label{eq:QNM_condition}
\end{equation}  
Here, $\Gamma_e$ denotes the phase contribution from a contour encircling the outer horizon. Since the horizon is a simple pole of $Q$, this contribution is determined by its residue 
\begin{equation}
    \Gamma_e = -2\pi \ii\  \underset{r \to r_+}{\text{Res}} \ Q = -\frac{\ii \, \omega}{2 T_{\rm H}}\ ,
    \label{eq:Gammae_Residue}
\end{equation}
where $T_{\rm H} = \sqrt{F'(r_+) \, B'(r_+)}/4\pi$ is the Hawking temperature of the outer horizon. $\Gamma_i$, instead, denotes the phase contribution associated with the closed contour $\mathcal{C}_i$ which encloses all singularities $r_{\rm p}$ other than the outer horizon (see inset in \cref{fig:SketchContour}). Its contribution is given by
\begin{equation}
    \Gamma_i = \oint_{\mathcal{C}_i} Q \, \dd r = -2\pi \ii \sum_{r_{\rm p} \in \text{Int}\left(\mathcal{C}_i \right)} \underset{r \to r_{\rm p}}{\text{Res}} \ Q\, .
\label{eq:Gamma_i_def}
\end{equation}
Finally, $\gamma \equiv |\gamma_{t_k}(t_{k+1})|$ denotes the phase integral between two consecutive turning points, $t_k$ and $t_{k+1}$.
As shown in Supplemental Material~\ref{app:Integral_phase},  
this phase is
\begin{equation}
    \gamma=\begin{cases}
        \pi/2 & r_0\ne 0 \\
        \frac{|N-1|}{N+1} \frac{\pi}{2}& r_0=0 
    \end{cases}\,,
\label{eq:gamma_evaluation}
\end{equation}
and its value remarkably depends on whether or not there is a singularity at $r=0$.

\begin{figure}
    \centering
    \includegraphics[width=\linewidth]{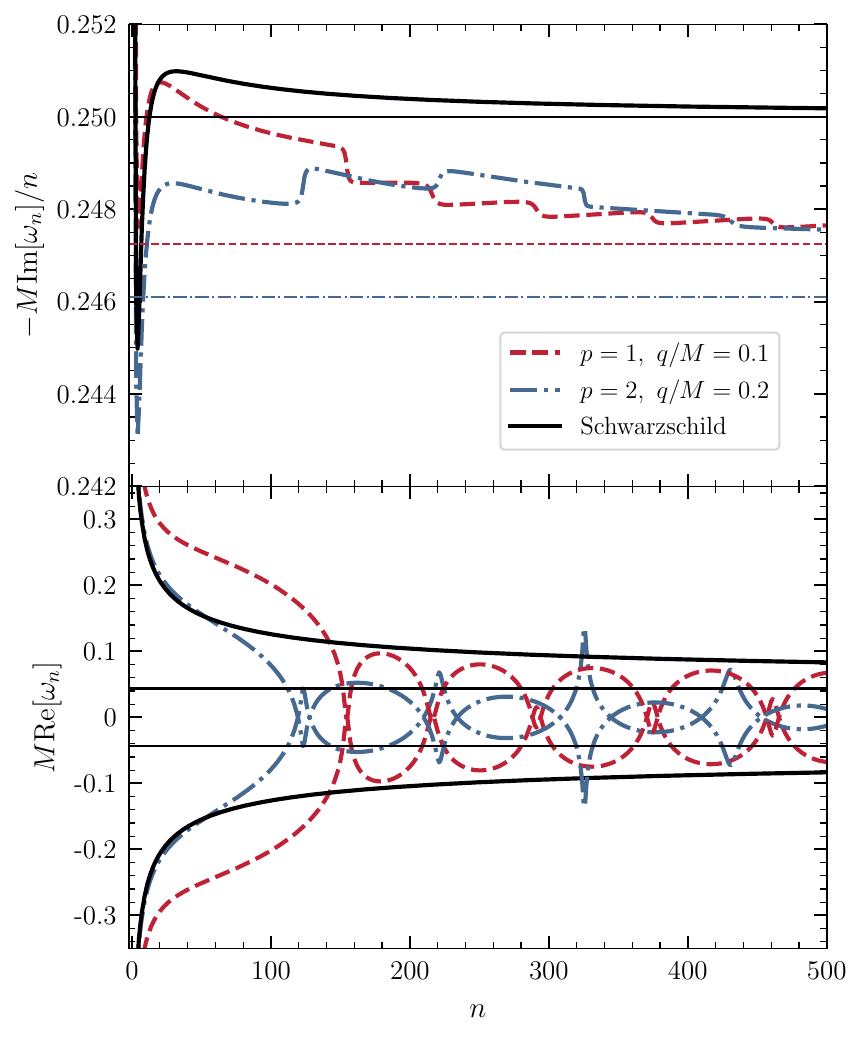}
    \caption{Asymptotic QNM spectra of regular BHs compared with Schwarzschild for the $\ell=2$ harmonic. Dashed red and dot-dashed blue curves show the first 500 overtones of the Fan--Wang BH ($p=1$, $q=0.1\,M$) and Bardeen BH ($p=2$, $q=0.2\,M$), respectively. Solid black curves show the Schwarzschild spectrum~\cite{Konoplya_qnms_tab}. Top and bottom panels display $-M\,\Im[\omega_n]/n$ and $M\,\Re[\omega_n]$, respectively. 
    Horizontal lines mark the predicted asymptotic limits: $-M\,\Im[\omega_n]/n\to 0.247$ (Fan--Wang), $0.246$ (Bardeen), and $1/4$ (Schwarzschild), while the real part vanishes for regular BHs, unlike the Schwarzschild limit $\ln 3/(8\pi)$ [see Eqs.~\eqref{eq:SchwarzschildQNMAsymptotic} and~\eqref{eq:UniversalBehavior_RegularBHs}]. 
    In the bottom panel we show both positive and negative frequencies to highlight the oscillations around zero. 
    }
    \label{fig:NumericalResults}
\end{figure}

%--------------------------------------------------------------------------------
{\bf\em Regular BHs.}
%--------------------------------------------------------------------------------
In this case, $r_0 \neq 0$ and, for all solutions satisfying the near-pole behavior in Eq.~\eqref{eq:metric_expansion_r0}, we have $\gamma=\pi/2$, independently of $F$ and $B$. Therefore, Eq.~\eqref{eq:QNM_condition} simplifies to
\begin{equation}
    \ee^{2\ii\Gamma_e}=1\,.
    \label{eq:QNM_cond_regular_BHs}
\end{equation}
Together with \cref{eq:Gammae_Residue}, this yields
\begin{equation}
    \Re[\omega_n]\to0\,,\qquad
    \Im[\omega_n]\to-2\pi T_{\rm H}\,n\,,
    \label{eq:UniversalBehavior_RegularBHs}
\end{equation}
as $n\to\infty$. Thus, under the assumptions of our analysis, static, asymptotically flat regular BHs with nondegenerate horizons possess a universal asymptotic QNM spectrum, in sharp contrast with the Schwarzschild result in Eq.~\eqref{eq:SchwarzschildQNMAsymptotic}. This result applies, for instance, to well-known nonsingular BHs with de Sitter cores~\cite{Cadoni:2022chn}, and it recovers the asymptotic spectrum obtained for a specific model in Ref.~\cite{Nomura:2005dn}. \Cref{eq:UniversalBehavior_RegularBHs} also extends to more general families interpolating between de Sitter and Minkowski cores~\cite{Hassanabadi:2026wrv}. 
For concreteness, we consider the class of regular BH spacetimes described by 
\begin{equation}
    B(r)=F(r)=1-\frac{2Mr^2}{\left(r^p+q^p\right)^{3/p}}\,.
    \label{eq:metricfunctionregulargeneral}
\end{equation}
Here, $M$ is the BH ADM mass, $p>0$ ensures regularity at $r=0$, and $q$ is the regularization scale, with the Schwarzschild geometry recovered as $q\to0$. This class includes the Fan--Wang~\cite{Fan:2016hvf} (see also Refs.~\cite{Cadoni:2022vsn,Cadoni:2023nrm}), Bardeen~\cite{bardeen1968proceedings}, and Hayward~\cite{Hayward:2005gi} geometries, corresponding to $p=1,2,3$, respectively. For any $p>0$, the geometry approaches a de Sitter core, $F(r)\sim1-r^2/L_{\rm dS}^2$, with $L_{\rm dS}=q^{3/2}/\sqrt{2M}$, and is asymptotically Schwarzschild. It possesses two, one degenerate, or no horizons for $q<q_{\rm ext}$, $q=q_{\rm ext}$, or $q>q_{\rm ext}$, respectively, where $q_{\rm ext}=2^{(p+2)/p}27^{-1/p}M$. Consistently with our assumption of no degenerate horizons, we always focus on $q < q_{\rm ext}$.
The family in Eq.~\eqref{eq:metricfunctionregulargeneral} belongs to the class discussed above, with $r_0\neq0$ and $z=N=3/p$. We therefore predict the universal spectrum in Eq.~\eqref{eq:UniversalBehavior_RegularBHs} for \emph{arbitrarily small}, nonzero $q$.
We numerically confirm this behavior for two representative regular BH geometries, the Fan--Wang and Bardeen solutions. 
\Cref{fig:NumericalResults} shows the first 500 overtones of the $\ell=2$ harmonic, obtained with a high-depth, high-precision continued-fraction implementation (see Supplemental Material~\ref{app:Regular_black_holes} for details).
The top panel shows $-M\,\Im[\omega_n]/n$, which slowly approaches $1/4$ for Schwarzschild, compared to $0.247$ and $0.246$ for the Fan--Wang and Bardeen BHs, respectively. The bottom panel highlights the qualitatively different behavior of the real part, which vanishes asymptotically for regular BHs, in contrast to the nonzero Schwarzschild limit $\ln 3/(8\pi)$. The numerical results are consistent with these predictions, illustrating the discontinuous change in the asymptotic QNM spectrum for arbitrarily small but nonzero $q$. 
Interestingly, ${\rm Re}[\omega_n]$ approaches zero through damped oscillations as $n\to\infty$. Oscillations in ${\rm Re}[\omega_n]$ also occur in the Reissner--Nordstr\"om case~\cite{Daghigh:2024wcl}, where, however, the real part does not vanish asymptotically. Each time ${\rm Re}[\omega_n]$ approaches zero, ${\rm Im}[\omega_n]/n$ exhibits a step-like feature. We have confirmed that this behavior persists qualitatively for other values of $q$.
Finally, note that our result differs from Ref.~\cite{Lan:2022qbb}, whose approximations do not fully capture the analytic structure of regular BH geometries, erroneously leading to a nonuniversal asymptotic spectrum.
%----------------------------------------------------------------------------
%----------------------------------------------------------------------------

%--------------------------------------------------------------------------------
{\bf\em Singular BHs.}
%--------------------------------------------------------------------------------
%
Our general result also applies to singular BHs. Singularities generally persist in effective-field-theory (EFT) extensions of GR~\cite{Endlich:2017tqa,Cardoso:2018ptl,Cano:2020cao,Cano:2019ore,Pani:2011gy,Cano:2021myl}, and even beyond the EFT regime~\cite{Sotiriou:2014pfa}. Although the high-frequency limit lies beyond any finite EFT cutoff, a modified but unresolved singularity may still leave an imprint on the asymptotic spectrum.
For single-horizon singular BHs ($r_0=0$), including those surrounded by arbitrary matter~\cite{Medved:2003rga}, the contour $\mathcal{C}_i$ in Fig.~\ref{fig:SketchContour} is absent. Equation~\eqref{eq:QNM_condition} then formally applies with $\Gamma_i\to\infty$, yielding 
\begin{equation}
\begin{split}
\Re[\omega_n]&\to T_{\rm H}\ln\left[1+2\cos\left(\frac{N-1}{N+1}\pi\right)\right]\,,\\
\Im[\omega_n]&\to -2\pi T_{\rm H}\,n\,,
\end{split}
\label{eq:QNM_condition_single_horizon}
\end{equation}
as $n\to\infty$, where we used \cref{eq:gamma_evaluation} for $r_0=0$. This result agrees with Ref.~\cite{Das:2004db}. The real part vanishes asymptotically only for $N=(3+2j)/(1-2j)$, with $j\in\mathbb{Z}$. Since $N>0$, the only possibilities are $N=1/3$ and $N=3$, whereas the Schwarzschild result is uniquely recovered for $N=1$. This highlights the sensitivity of the asymptotic spectrum to the structure of the central singularity.

For singular BH geometries with multiple nondegenerate horizons, $\Gamma_i$ is given by Eq.~\eqref{eq:Gamma_i_def}. Equation~\eqref{eq:QNM_condition} then shows that a vanishing asymptotic real part requires $\gamma=\pi/2+j\pi$, with $j\in\mathbb{Z}$, which is incompatible with Eq.~\eqref{eq:gamma_evaluation} since $N>0$. Thus, within the large class considered here, an asymptotically purely damped spectrum is a distinctive signature of regular BHs, with the exceptional fine-tuned cases $N=1/3$ and $N=3$ for single-horizon singular BHs.

%%%%%%%%%%%%%%%%%%%%%%%%%%%
{\bf\em Connection to the BH spectral instability.}
%%%%%%%%%%%%%%%%%%%%%%%%%%%
The seemingly paradoxical noncommutativity of the limits $q\to0$ and $n\to\infty$ is naturally understood in terms of the BH spectral instability~\cite{Nollert:1996rf,Barausse:2014tra,Jaramillo:2020tuu,Cheung:2021bol,Berti:2022xfj,Cardoso:2024mrw,Miyachi:2025dyk}. Owing to the non-self-adjoint nature of the QNM spectral problem, arising from radiative boundary conditions at the horizon and at infinity, arbitrarily small perturbations of the background geometry can produce large spectral shifts. For highly localized perturbations, these shifts can grow exponentially with $|{\rm Im}[\omega_n]|$~\cite{Leung:1997was,Ianniccari:2024ysv}. Since $|{\rm Im}[\omega_n]|$ grows linearly with the overtone number $n$, any fixed nonzero deformation can eventually produce order-unity spectral changes at sufficiently large $n$. The large-overtone limit thus acts as a magnifying glass for corrections that are arbitrarily small in the exterior geometry.
Indeed, Eq.~\eqref{eq:QNM_condition_single_horizon} shows that a similar discontinuity occurs for any modification of the inner singularity, such as those that may arise in UV modifications of GR.
Although in this case, too, the GR limit does not commute with $n\to\infty$, the universal behavior in Eq.~\eqref{eq:UniversalBehavior_RegularBHs} remains a hallmark of regular BHs. Thus, while the noncommutativity of the two limits is naturally associated with spectral instability, the specific asymptotic behavior of the real part retains information about the nature of the underlying deformation.

%%%%%%%%%%%%%%%%%%%%%%%%%
{\bf \em Discussion.}
%%%%%%%%%%%%%%%%%%%%%%%%%
Unfortunately, the intriguing possibility of distinguishing a regular BH from a singular one through the asymptotic QNM spectrum appears extremely challenging observationally. Although the corrections in ${\rm Re}[\omega_n]$ are of order unity, 
the transition to the universal regular-BH spectrum sets in at very high overtone number. 
This transition occurs when the characteristic size of the turning-point clusters, $\delta r\equiv |t_k-r_0|$, becomes comparable to the displacement $|r_0|$ of the singularities from the origin. For the family~\eqref{eq:metricfunctionregulargeneral}, this can be estimated from the local WKB structure as (see Supplemental Material~\ref{app:estimate})
\begin{equation}
    \delta r\sim \frac{1}{p}\left[
\frac{M(3+p)q^{-1+3/p}}{|\omega|}
\right]^{p/(3+p)}.
\end{equation}
Requiring $\delta r\sim q$ gives a critical frequency,
\begin{equation}
    |\omega_c|\sim
    \frac{p+3}{p^{1+3/p}}\frac{M}{q^2}\,,
\end{equation}
and, since $|\omega_n|M\propto n$ at large $n$, a critical overtone number
\begin{equation}
    n_c\sim\frac{p+3}{p^{1+3/p}}\left(\frac{M}{q}\right)^2\,,
    \label{eq:critical_overtone}
\end{equation}
up to a model-dependent coefficient of order unity. Remarkably, the scaling with the regularization parameter is independent of $p$. For a Planckian regularization scale, $q\sim\ell_{\rm P}$, $n_c$ is therefore enormous for any astrophysical BH. Even for $q\sim0.01M$, one expects $n_c\sim\mathcal{O}(10^4)$.

There is a second, closely related observational obstacle. Highly damped overtones are, by definition, increasingly short lived: since their damping time scales as $\tau_n\sim1/|{\rm Im}[\omega_n]|$, and
$|{\rm Im}[\omega_n]|$ grows approximately linearly with $n$, their lifetime decreases roughly as $1/n$. They therefore decay on progressively shorter timescales and are increasingly difficult to disentangle from the prompt response and from neighboring overtones. Given the challenges already involved in robustly extracting the first few overtones from gravitational-wave
data~\cite{Baibhav:2023clw,Cheung:2023vki,Zhu:2023mzv,Mitman:2025hgy,Berti:2025hly}, accessing $n\sim{\cal O}(10^4)$---let alone the vastly larger values associated with a putative microscopic regularization scale---appears out of reach even for future detectors.

An interesting open question is whether the nonperturbative sensitivity of highly damped modes to singularity resolution can overcome their limited observability and yield stronger constraints than the more precisely measurable, but parametrically smaller ${\cal O}[(q/M)^p]$ deviations of the fundamental mode.

Furthermore, the asymptotic QNM spectrum may leave an imprint on the global ringdown response, beyond the direct observation of individual highly damped overtones. For example, recent work~\cite{Arnaudo:2026tcy} suggests that the asymptotic spectrum may affect the analytic structure of the full retarded Green's function and, consequently, the ringdown signal.

The significance of our result is primarily theoretical. It provides a sharp example in which an arbitrarily small modification of the geometry is amplified in the large-overtone limit, making the limits $q\to0$ and $n\to\infty$ noncommuting. At the same time, the specific asymptotic spectrum distinguishes singularity resolution from generic high-curvature corrections, establishing a new connection between BH spectral instability and the analytic structure of regular BH geometries.

Although we have focused on test scalar perturbations of spherically symmetric BHs, we expect the qualitative features of our results to persist for gravitational perturbations and spinning BHs. Another natural extension concerns degenerate horizons, for which the analytic structure underlying the monodromy calculation changes~\cite{Medved:2003rg,Daghigh:2007xj} and the interplay between the large-overtone and extremal limits may lead to qualitatively different asymptotic spectra.

Finally, since ${\rm Im}[\omega_n]$ retains the spacing $2\pi T_{\rm H}$, Maggiore's area quantum $\Delta A=8\pi\ell_{\rm P}^2$~\cite{Maggiore:2007nq} is unaffected.

% Finally, the spectral discontinuity found here does not affect Maggiore's proposal for BH area quantization~\cite{Maggiore:2007nq}: the proper frequency is asymptotically dominated by $|{\rm Im}[\omega_n]|$, whose spacing remains $2\pi T_{\rm H}$ also for regular BHs, thus yielding the standard area quantum $\Delta A=8\pi\ell_{\rm P}^2$.

\noindent
{\bf \em  Acknowledgments.}
We thank L.~Capuano and M.~Minucci for useful discussions. We acknowledge financial support by the MUR FIS2 Advanced Grant ET-NOW (CUP:~B53C25001080001) and by the INFN TEONGRAV initiative. 
Some numerical computations have been performed at the Vera and CHRONOS clusters supported by the Italian Ministry of Research and by Sapienza University of Rome.
%

%----------------------------------------------------------------------------

\bibliographystyle{apsrev4-1}
\bibliography{Refs.bib}

\onecolumngrid
\clearpage
%----------------------------------------------------
\section*{Supplemental Material}
%----------------------------------------------------------------------------
\renewcommand{\thesubsection}{{SM.\arabic{subsection}}}
\setcounter{section}{0}
%----------------------------------------------------------------------------
%----------------------------------------------------------------------------
\subsection{Definition of \texorpdfstring{$Q$}{Q} and evaluation of \texorpdfstring{$\gamma$}{gamma}}\label{app:Integral_phase}
%----------------------------------------------------------------------------
Inspired by the usual WKB ansatz, we initially set $Q^2=R$.
In the high-frequency limit, $|\omega|\to\infty$, the turning points, defined as the roots of $Q^2$, accumulate near the poles of $\sqrt{H}$. The phase $\gamma_{t_k}(t_{k+1})$ between two consecutive turning points, $t_k$ and $t_{k+1}$, can therefore be evaluated from the local behavior of $Q$ around these poles. Let $r_0\in\mathbb{C}$ be such a pole, and assume that the metric functions behave locally as
\begin{equation}
    F(r)=(r-r_0)^{-2N+z}f(r)\,,\qquad
    B(r)=(r-r_0)^{-z}b(r)\,,
    \label{eq:metric_r0}
\end{equation}
where $f(r)$ and $b(r)$ are regular and nonvanishing at $r_0$, and $N>0$.

Substituting \cref{eq:metric_r0} into \cref{eq:R} and retaining the leading terms as $|\omega|\to\infty$ and $r\to r_0$, we obtain
\begin{equation}
    R \sim
    \frac{\omega^2}{b(r_0)f(r_0)}(r-r_0)^{2N}
    -\frac{N(N\pm2)}{4(r-r_0)^2}
    -\frac{\ell(\ell+1)}{b(r_0)}
    \frac{(r-r_0)^z}{r^2}\,,
    \label{eq:general_expansion_at_r0}
\end{equation}
with $+$ for $r_0\neq0$ and $-$ for $r_0=0$. The last term requires some care. For $r_0\neq0$, it scales as $(r-r_0)^z$ and is subleading with respect to the inverse-square term for $z>-2$. For $r_0=0$, instead, it scales as $r^{z-2}$ and is subleading for $z>0$. We first consider these generic cases, in which the local behavior is controlled by the inverse-square term. The limiting cases $z=-2$ for $r_0\neq0$ and $z=0$ for $r_0=0$, the special case $r_0=0$ with $N=2$, and smaller values of $z$ are discussed below.

The standard choice $Q^2=R$ does not reproduce the correct local behavior of the solution near $r_0$~\cite{Langer:1937qr,Andersson:2003fh}. Indeed, sufficiently close to $r_0$, where the frequency-dependent term in \cref{eq:general_expansion_at_r0} is subleading, \cref{eq:EqPerturbationGeneral} reduces to
\begin{equation}
    \frac{\dd^2\Psi}{\dd r^2}
    -\frac{\alpha}{(r-r_0)^2}\Psi=0\,,
    \qquad
    \alpha\equiv
    \frac{N(N\pm2)}{4}\,.
    \label{eq:perteq_at_r0}
\end{equation}
Matching the WKB solutions~\eqref{eq:WKBgeneralSolution} to the local solutions of \cref{eq:perteq_at_r0} requires
\begin{equation}
    Q^2\sim-\frac{1+4\alpha}{4(r-r_0)^2}\,.
    \label{eq:Q_exp_r0}
\end{equation}
Indeed, this behavior ensures that the WKB solutions $f^{(t)}_{1,2}$ reproduce the two local power-law solutions of \cref{eq:perteq_at_r0}.

In the complementary region, where the frequency-dependent term dominates, the perturbation equation reduces to
\begin{equation}
    \frac{\dd^2\Psi}{\dd r^2}
    +\mathcal{P}^2\omega^2(r-r_0)^{2N}\Psi=0\,,
    \qquad
    \mathcal{P}\equiv
    \left[b(r_0)f(r_0)\right]^{-1/2}\,,
\label{eq:perteq_at_r0_omega}
\end{equation}
which yields
\begin{equation}
    Q^2\sim
    \mathcal{P}^2\omega^2(r-r_0)^{2N}\,.
    \label{eq:Q_r0_omega}
\end{equation}
Combining \cref{eq:Q_exp_r0,eq:Q_r0_omega}, we obtain the local WKB function that interpolates between the two regimes,
\begin{equation}
    Q\sim
    \pm\mathcal{P}\omega(r-r_0)^N
    \left[
    1-\frac{1+4\alpha}
    {4\mathcal{P}^2\omega^2(r-r_0)^{2(N+1)}}
    \right]^{1/2}.
    \label{eq:Q_exp_general_form}
\end{equation}
The two signs correspond to the two square-root branches. This expression is equivalent, at the relevant order, to the Langer-type prescription~\cite{Langer:1937qr,Andersson:2003fh}
\begin{equation}
    Q^2=R-\frac{1}{4(r-r_0)^2}\,,
    \label{eq:Q_Langer}
\end{equation}
which therefore reproduces the correct local behavior near $r_0$.

We can now evaluate the phase integral between consecutive turning points. Substituting \cref{eq:Q_exp_general_form} into $\gamma_t(r)$ [i.e., Eq.~\eqref{eq:WKBgeneralSolution}] and introducing
\begin{equation}
    y\equiv
    \frac{2\omega\mathcal{P}}{\sqrt{1+4\alpha}}
    (r-r_0)^{N+1}\,,
    \label{eq:yr_Map}
\end{equation}
we obtain
\begin{equation}
    \gamma
    \equiv \left|\gamma_{t_k}(t_{k+1})\right|
    =\frac{\sqrt{1+4\alpha}}{2(N+1)}
    \left|
    \int_{-1}^{1}\dd y\,\sqrt{1-y^{-2}}
    \right|
    =\frac{\sqrt{1+4\alpha}}{2(N+1)}\pi\,.
    \label{eq:intI}
\end{equation}
The transformation~\eqref{eq:yr_Map} maps consecutive roots $t_k$ of $Q^2$ alternately to $y=\pm1$. More precisely, $y_k=(-1)^{k+j}$, where $j=\{0,1\}$ labels the square-root branch. Fixing $y_1=-1$ determines the mapping of all remaining roots and hence the sign of $\gamma_{t_k}(t_{k+1})$.

For $r_0\neq0$ and $z>-2$, one has 
\begin{equation}
    \alpha=\frac{N(N+2)}{4}\,,
    \qquad
    \gamma=\frac{\pi}{2}\,.
\end{equation}
For $r_0=0$ and $z>0$, instead,
\begin{equation}
    \alpha=\frac{N(N-2)}{4}\,,
    \qquad
    \gamma=\frac{|N-1|}{2(N+1)}\pi\,.
\end{equation}
These results yield \cref{eq:gamma_evaluation} in the main text. Note that this applies whenever $F=B$, as $z=N>0$, except for $r=0$ and $N=2$.
We finally discuss the limiting cases in which the angular-momentum term contributes at the same order as the geometrical inverse-square term. For $r_0\neq0$ and $z=-2$, \cref{eq:perteq_at_r0} still applies, but with
\begin{equation}
    \alpha=
    \frac{N(N+2)}{4}
    +\frac{\ell(\ell+1)}{b(r_0)r_0^2}\,.
    \label{eq:alpha_r0_nonzero_boundary}
\end{equation}
Similarly, for $r_0=0$ and $z=0$,
\begin{equation}
    \alpha=
    \frac{N(N-2)}{4}
    +\frac{\ell(\ell+1)}{b(0)}\,.
    \label{eq:alpha_r0_zero_boundary}
\end{equation}
In both cases, the derivation above remains valid upon replacing $\alpha$ by the corresponding effective coefficient, but $\gamma$ becomes explicitly $\ell$-dependent through \cref{eq:intI}.
A particularly interesting example is $r_0=0$, $N=2$, and $z=0$, for which the geometrical contribution vanishes and
\begin{equation}
    \alpha=\frac{\ell(\ell+1)}{b(0)}\,.
\end{equation}
This case is relevant to certain loop-quantum-gravity-inspired singular BHs~\cite{Modesto:2008im}. Substituting this value of $\alpha$ into \cref{eq:intI} and using the monodromy condition~\eqref{eq:QNM_condition}, we recover the results of Ref.~\cite{Montagnon:2025kyn}.
For $r_0=0$, $N=2$, and $z\geq 2$, the function $R$ in \cref{eq:general_expansion_at_r0} is instead regular at $r=0$. In this case, the leading $\omega^2$-dependent term dominates in the large-$|\omega|$ limit also at $r\to0$. Consequently, the phase $\gamma$ cannot be determined using the local WKB analysis adopted above. 
Nevertheless, retaining the leading term reduces the Schr\"odinger-like equation to \cref{eq:perteq_at_r0_omega} with $N=2$, whose general solution is 
\begin{equation}
  \Psi=   \sqrt{r \omega} \left[c_1 J_{-\frac{1}{6}}\left(\frac{1}{3} r^3 \mathcal{P} \omega \right)+c_2
    J_{\frac{1}{6}}\left(\frac{1}{3} r^3 \mathcal{P} \omega \right) \right]\,,
\label{eq:psi_N_2_reg}
\end{equation}
where $J_{\pm 1/6}$ are Bessel functions and $c_1$ and $c_2$ are integration constants.
The large-argument asymptotics of \cref{eq:psi_N_2_reg} describe superpositions of ingoing and outgoing waves, whose relative amplitudes change upon analytic continuation in the complex-$r$ plane. The asymptotic QNM spectrum can therefore be obtained through the matching procedure of Ref.~\cite{Motl:2003cd}. For BH geometries admitting a contour analogous to $\Lambda$ in \cref{fig:SketchContour}, the resulting spectrum is given by Eq.~(62) of that reference, with $j=1/3$ and $\delta=\Gamma_i$, defined in Eq.~\eqref{eq:Gamma_i_def}. This class includes the Reissner--Nordstr\"om solution, corresponding to $z=2$.
Finally, for $r_0\neq0$ with $z<-2$, or $r_0=0$ with $z<0$, the angular-momentum term is more singular than the inverse-square contribution. 
The local structure then differs from \cref{eq:perteq_at_r0}, and these cases fall outside the class considered here.
%
%----------------------------------------------------------------
\subsection{Anti-Stokes network and existence of the monodromy contour}
\label{app:path_existence}
%----------------------------------------------------------------------------
We argue that, for a broad class of static, spherically symmetric, asymptotically flat BH geometries with two nondegenerate horizons and Schwarzschild asymptotics, the highly damped regime admits an anti-Stokes contour analogous to that shown in \cref{fig:SketchContour} of the main text. This contour extends from spatial infinity, successively encircles the outer and inner horizons, and returns to its starting point. Our argument relies on the structure of the anti-Stokes network near infinity, around the horizons, and in the vicinity of the singularities of $FB$, together with the constraint that distinct anti-Stokes lines cannot intersect or terminate at regular points away from turning points.
The leading-order contribution to $Q^2$ in the highly damped regime,  $|\omega|M\gg1$,
\begin{equation}
    Q^2 \sim \frac{\omega^2}{F \, B}\, ,
\end{equation}
is sufficient for the analysis below.
By assumption, the geometries under consideration approach Schwarzschild at spatial infinity. Thus, the asymptotic anti-Stokes lines coincide with those of Schwarzschild and satisfy
\begin{equation}
    \text{Im}\left[\omega \left(\bar r + \ln\left(\bar r-1 \right) \right)  \right] = 0\, , \qquad \bar r \equiv \frac{r}{2M}\, .
\end{equation}
For large, purely imaginary frequencies, $\omega=-\ii|\omega|$, this condition reduces to $\mathrm{Re}(\bar r)+\ln|\bar r-1|=0$, yielding two asymptotic directions in the complex $r$-plane, $\arg(\bar r)\to\pm\pi/2$.
Near the two nondegenerate horizons $r_\pm$, which are simple zeros of both $F$ and $B$, we have
\begin{equation}
\int Q\,\dd r \approx \frac{\omega}{\sqrt{F'(r_\pm)B'(r_\pm)}}\ln(r-r_\pm)\,.
\end{equation}
Writing $r-r_\pm=\rho \,\ee^{\ii\varphi}$, the anti-Stokes condition for purely imaginary $\omega$ fixes $\rho$ to a constant. Thus, the anti-Stokes lines are locally circular around each horizon, constraining the possible global connectivity of the network.

The global connectivity of the anti-Stokes network is further
constrained by its local structure near the singularities of $FB$.
Consider a pole $r_0$ of order $2N$, where
$FB\sim f(r_0)b(r_0)(r-r_0)^{-2N}$, following the notation of~\ref{app:Integral_phase}.
Near $r_0$, the anti-Stokes lines satisfy
\begin{equation}
    \text{Im}\left[\int Q\,\dd r\right]
    = \text{Re}\left[
    -\frac{|\omega|}{\sqrt{f(r_0)b(r_0)}}
    \frac{(r-r_0)^{N+1}}{N+1}
    \right]=0\,.
\end{equation}
These lines may extend to infinity or connect to other regions of the network, including those surrounding the horizons. Since $\int Q\,\dd r$ is locally invertible wherever $Q$ is analytic and nonzero, distinct anti-Stokes lines cannot intersect at regular points, but only at turning points where $Q=0$. 
Branch points require special care, as the network must be consistently tracked across the relevant Riemann surface.
Disconnected components may also extend to infinity without terminating at turning points. Nevertheless, for geometries approaching Schwarzschild asymptotically, the network retains the Schwarzschild structure at large distances, while its behavior near the horizons constrains the global connectivity. We therefore expect a contour $\Lambda$ connecting the asymptotic region to the horizons and encircling both horizons to exist, even in the presence of disconnected components and branch cuts.

We explicitly verify the existence of such a contour in
Supplemental Material~\ref{app:Regular_black_holes} for the
Fan--Wang and Bardeen BHs, whose analytic structures are
respectively free of and exhibit branch points.

%----------------------------------------------------------------------------
\subsection{Derivation of the general monodromy condition~\eqref{eq:QNM_condition}}
\label{subsubsec:DerivationMonodromyGeneralContour}

As the solution $\Psi$ of \cref{eq:EqPerturbationGeneral} is
analytically continued along the contour $\Lambda$ shown in \cref{fig:SketchContour}, its WKB coefficients evolve according to the Stokes and anti-Stokes structure. Upon crossing a Stokes line, the coefficient of the dominant solution remains unchanged, while that of the subdominant solution acquires a contribution
proportional to the former. This is the Stokes phenomenon~\cite{Berry:1972na,Berry:1989uni,Andersson:2003fh}.
Across an anti-Stokes line, the roles of the dominant and subdominant solutions interchange (see Ref.~\cite{Andersson:2003fh} for details).

Analytic continuation between adjacent turning points introduces phase factors determined by the phase integral $\gamma$, evaluated
in~\ref{app:Integral_phase}.
The signs of these phases depend on the mapping of the turning points to $\pm1$ through \cref{eq:yr_Map}. We adopt the convention
$t_1\mapsto-1$, where $t_1$ is the turning point connected to infinity by an anti-Stokes line (see \cref{fig:SketchContour}),
and consequently map successive turning points alternately to $+1$ and $-1$ in the clockwise direction. Defining the phase accumulated along $-1\to+1$ as $-\gamma$, we obtain
\begin{equation}
    \gamma_{12} = \gamma_{32} = -\gamma\, ,
    \label{eq:gamma1232general}
\end{equation}
where we introduced the shorthand notation $\gamma_{ij} \equiv \gamma_{t_i}(t_j)$. 
The sign of the phase accumulated from $t_{i+1}$ to $t_i$
depends on the parity of $i$:
\begin{equation}
    \gamma_{i+1,\,i}\equiv\bar\gamma=(-1)^{i+1}\gamma\,.
    \label{eq:barGamma}
\end{equation}
We also need the phases accumulated along the paths
encircling the outer and inner horizons, denoted by
$\tilde\gamma_{2,\,i+1}$ and $\tilde\gamma_{i,\,3}$,
respectively.
The phase $\Gamma_e$ associated with a closed contour
enclosing only the outer horizon can be decomposed as
(see \cref{fig:SketchContour})
\begin{equation}
    \Gamma_e=
    \tilde\gamma_{2,\,i+1}
    +\bar\gamma
    +\tilde\gamma_{i,\,3}
    +\gamma_{32}\,.
\end{equation}
Using \cref{eq:gamma1232general}, we obtain
\begin{equation}
    \tilde\gamma_{2,\,i+1}
    =\Gamma_e-\bar\gamma-\tilde\gamma_{i,\,3}+\gamma\,.
    \label{eq:Gammae_General}
\end{equation}
We now analytically continue $\Psi$ along the contour $\Lambda \equiv a \to b \to c \to c' \to d \to e \to e' \to f \to c \to b \to a$ (see \cref{fig:SketchContour}). 

We start from region $a$, denoting asymptotic infinity. With the chosen branch of $Q$, the outgoing solution is proportional to $f_1$. As no Stokes lines are crossed
\begin{equation}
    \Psi_a = f_1^{(t_1)}\, .
\end{equation}
Analytic continuation to point $b$ implies traversing a Stokes line in the clockwise direction. In this region, $f_1$ dominates over $f_2$, so taking into account the Stokes phenomenon~\cite{Andersson:2003fh}, we obtain
\begin{equation}
    \Psi_b = f_1^{(t_1)} - \ii f_2^{(t_1)}\, .
\end{equation}
In moving to point $t_2$, both $f_1$ and $f_2$ acquire a phase $\gamma_{12}$
\begin{equation}
    \Psi_b = \ee^{\ii \gamma_{12}} f_1^{(t_2)} - \ii \ee^{-\ii \gamma_{12}} f_2^{(t_2)} = \ee^{-\ii \gamma} f_1^{(t_2)} - \ii \ee^{\ii \gamma}f_2^{(t_2)}\, ,
\end{equation}
where we also used \cref{eq:gamma1232general}. Moving to region $c$, another Stokes line is crossed. Since $f_1$ is still the dominant solution, we have
\begin{equation}
    \Psi_c = \ee^{-\ii \gamma} f_1^{(t_2)} - \ii \left(\ee^{\ii \gamma} + \ee^{-\ii \gamma} \right)f_2^{(t_2)}\, .
\end{equation}
We next move to the point $c'$ along the contour encircling the outer horizon. Both $f_1$ and $f_2$ will acquire a phase $\tilde \gamma_{2, \, i+1}$. Using \cref{eq:Gammae_General}, we obtain
\begin{equation}
    \Psi_{c'} = \ee^{\ii \left(\Gamma_e - \bar \gamma - \tilde \gamma_{i, \, 3} \right)}f_1^{(t_{i+1})} - \ii \ee^{-\ii \left(\Gamma_e - \bar \gamma - \tilde \gamma_{i, \, 3} \right)} \left(1+\ee^{-2\ii \gamma} \right) f_2^{(t_{i+1})} \, .
\end{equation}
Analytically continuing to region $d$ implies crossing first the anti-Stokes line of the contour around $r_+$. This means that inside the big loop, $f_2$ dominates over $f_1$. Additionally, we have to cross a Stokes line in anticlockwise direction. Taking into account the Stokes phenomenon and also moving to $t_i$ yields
\begin{equation}
    \Psi_d = \left[\ee^{\ii \left(\Gamma_e - \tilde \gamma_{i, \, 3} \right)} + \ee^{-\ii \left(\Gamma_e - 2 \bar \gamma - \tilde \gamma_{i, \, 3} \right)} \left(1 + \ee^{-2\ii \gamma} \right) \right] f_1^{(t_i)} - \ii \ee^{-\ii \left(\Gamma_e - \tilde \gamma_{i, \, 3} \right)}\left(1 + \ee^{-2\ii \gamma} \right) f_2^{(t_i)}\, .
\end{equation}
Traversing another Stokes line we arrive in region $e$, where
\begin{equation}
    \Psi_e = \left[\ee^{\ii \left(\Gamma_e - \tilde \gamma_{i, \, 3} \right)} + \ee^{-\ii \left(\Gamma_e - \tilde \gamma_{i, \, 3} \right)} \left(1 + \ee^{-2\ii \gamma} \right) + \ee^{-\ii \left(\Gamma_e - 2 \bar \gamma - \tilde \gamma_{i, \, 3} \right)} \left(1 + \ee^{-2\ii \gamma} \right) \right]f_1^{(t_i)} - \ii \ee^{-\ii \left(\Gamma_e - \tilde \gamma_{i, \, 3} \right)} \left(1 + \ee^{-2\ii \gamma} \right) f_2^{(t_i)}\, .
\end{equation}
We now continue to region $e'$ along the contour encircling the inner horizon. Both $f_1$ and $f_2$ will acquire a phase proportional to $\tilde \gamma_{i, \, 3}$, giving 
\begin{equation}
    \Psi_{e'} = \left[\ee^{\ii \Gamma_e} + \ee^{-\ii \left(\Gamma_e - 2 \tilde \gamma_{i, \, 3} \right)} \left(1 + \ee^{-2\ii \gamma} \right) + \ee^{-\ii \left(\Gamma_e - 2 \bar \gamma - 2 \tilde \gamma_{i, \, 3} \right)} \left(1 + \ee^{-2\ii \gamma} \right)\right]f_1^{(t_3)} - \ii \ee^{-\ii \Gamma_e} \left(1 + \ee^{-2\ii \gamma} \right) f_2^{(t_3)}\, .
\end{equation}
Moving to region $f$, we cross another Stokes line and we analytically continue the solution to $t_2$, obtaining, using \cref{eq:gamma1232general},
\begin{equation}
\begin{split}
    \Psi_f = &\left[\ee^{\ii \left(\Gamma_e - \gamma \right)} + \ee^{-\ii \left(\Gamma_e + \gamma \right)} \left(1 + \ee^{-2\ii \gamma} \right) + \ee^{-\ii \left(\Gamma_e - 2 \tilde \gamma_{i, \, 3} + \gamma \right)}\left(1 + \ee^{-2\ii \gamma} \right) + \ee^{-\ii \left(\Gamma_e - 2 \bar \gamma - 2 \tilde \gamma_{i, \, 3} + \gamma \right) }\left(1 + \ee^{-2\ii \gamma} \right)\right]f_1^{(t_2)}\\
    &- \ii \ee^{-\ii \left(\Gamma_e - \gamma \right)}\left(1 + \ee^{-2\ii \gamma} \right)f_2^{(t_2)}\, . 
\end{split}
\end{equation}
We move again to region $c$, crossing another Stokes line (a bar over the subscript denotes that we are traversing the corresponding region again after completing the loops around $r_\pm$):
\begin{equation}
    \Psi_{\bar c} = \mathcal{A} \, \ee^{-\ii\gamma} f_1^{(t_2)} - \ii \ee^{-\ii \left(\Gamma_e - \gamma \right)}\left(1 + \ee^{-2\ii \gamma} \right) f_2^{(t_2)}\, ,
\end{equation}
where we defined 
\begin{equation}
    \mathcal{A}  \equiv \ee^{\ii \Gamma_e } + \left(1+ \ee^{-2\ii \gamma} \right) \biggl[\ee^{-\ii \Gamma_e} + \ee^{-\ii \left(\Gamma_e - 2 \tilde \gamma_{i, \, 3} \right)} + \ee^{-\ii \left(\Gamma_e- 2 \bar \gamma - 2 \tilde \gamma_{i, \, 3}  \right)} + \ee^{-\ii \left(\Gamma_e -2 \gamma \right)} \biggr] \, .
    \label{eq:A_Monodromy_General}
\end{equation}
Returning to region $b$, we cross a Stokes line. Here, $f_1$ is again dominant over $f_2$, so the Stokes phenomenon and continuing the solution to $t_1$ give
\begin{equation}
    \Psi_{\bar b} = \mathcal{A} \, f_1^{(t_1)} - \ii \left[-\mathcal{A} \, \ee^{-2\ii \gamma} + \ee^{-\ii \Gamma_e} \left(1 + \ee^{-2\ii \gamma} \right) \right]f_2^{(t_1)}\, .
\end{equation}
Continuing the solution to the starting point $a$ yields, after traversing the Stokes line, 
\begin{equation}
    \Psi_{\bar a} = \mathcal{A}\,   f_1^{(t_1)}+\mathcal{B}f_2^{(t_1)}\, , \qquad \mathcal{B}\equiv  - \ii \left(1 + \ee^{-2\ii \gamma} \right) \left(\ee^{-\ii \Gamma_e}-\mathcal{A} \right) \, .
\end{equation}
Since a QNM eigenfunction $\Psi=\Psi_{\rm QNM}$ must be an eigenfunction of the monodromy transformation~\cite{Motl:2003cd}, the QNM condition is $\mathcal{B}=0$, namely
\begin{equation}
    \left(1 + \ee^{-2\ii \gamma} \right) \left(\ee^{-\ii \Gamma_e}-\mathcal{A} \right) = 0\, .
    \label{eq:MonodromyConditions0}
\end{equation}
For general values of $\gamma$, the factor $1+\ee^{-2\ii\gamma}$ is nonzero and
\cref{eq:MonodromyConditions0} reduces to $\ee^{-\ii \Gamma_e} =\mathcal{A}$. Using \cref{eq:A_Monodromy_General}, we obtain
\begin{equation}
    \ee^{2\ii \Gamma_e} = 1 - \left(1 + \ee^{-2\ii \gamma} \right) \left[1 + \ee^{2\ii \gamma} + \ee^{2\ii \tilde \gamma_{i, \, 3}} \left(1 + \ee^{2\ii \bar \gamma} \right) \right]\, .
    \label{eq:FinalResult_Monodromy_General_SM_Temp}
\end{equation}
Note that for regular BHs, $\gamma=\pi/2$ at leading WKB order [see \cref{eq:gamma_evaluation}], seemingly reducing Eq.~\eqref{eq:MonodromyConditions0} to a trivial identity. However, subleading corrections generically shift $\gamma$ away from $\pi/2$, allowing the QNM condition~\eqref{eq:FinalResult_Monodromy_General_SM_Temp} to be recovered at leading order.
We now evaluate $\tilde \gamma_{i, \,3}$, which depends on the arrangement of turning points and poles enclosed by the smaller loop surrounding $r_-$.
Let $\mathcal{N}$ denote the number of intermediate turning points along the chain connecting $t_3$ and $t_i$, as illustrated in the inset of \cref{fig:SketchContour}.
The internal closed contour $\mathcal{C}_i$ around the inner horizon can then be decomposed into the loop connecting $t_3$ and $t_i$, and a chain of $\mathcal{N}+1$ segments joining the internal turning points $\left(t_3\, , \dots, \, t_i\right)$.  We denote by $\Gamma_i$ the phase accumulated along $\mathcal{C}_i$ and note that the phase contribution alternates between $\pm\gamma$ along successive segments, i.e., $\gamma_{34} = -\gamma$, $\gamma_{45} = \gamma$ and so on, according to the conventions outlined at the beginning of the section. $\Gamma_i$ can then be written as
\begin{equation}
    \Gamma_i = -\tilde \gamma_{i, \, 3} - \sum_{\lambda = 1}^{\mathcal{N}+1} (-1)^{\lambda} \gamma\, .
    \label{eq:GammaI_General}
\end{equation}
The alternating sum equals $-\gamma$ (vanishes) for even (odd) $\mathcal N$. Since $i = \mathcal{N}+4$ (see inset in \cref{fig:SketchContour}), we have, using also \cref{eq:barGamma}
\begin{equation}
\begin{aligned}
\mathcal N \text{ even}: \quad
&\tilde\gamma_{i,\,3}=-\Gamma_i+\gamma,
\qquad &&\bar\gamma=-\gamma\,,
\\
\mathcal N \text{ odd}: \quad
&\tilde\gamma_{i,\,3}=-\Gamma_i,
&&\bar\gamma=\gamma.
\end{aligned}
\end{equation}
Substituting either pair of expressions into \cref{eq:FinalResult_Monodromy_General_SM_Temp} always yields \cref{eq:QNM_condition}. 
Thus, the monodromy condition takes a universal form, independent of the number and arrangement of turning points enclosed by the inner contour. All model dependence is encoded in the phase integral $\gamma$ and the monodromies $\Gamma_e$ and $\Gamma_i$, the latter determined by the residues of $Q$ at the enclosed singularities.
 %--------------------------------------------------------
\subsection{Quasinormal modes of Fan--Wang and Bardeen black holes}
\label{app:Regular_black_holes}
%--------------------------------------------------------
We test the general prediction~\eqref{eq:wR} for two representative
regular BHs belonging to the family~\eqref{eq:metricfunctionregulargeneral}.
We first establish the existence of a monodromy contour $\Lambda$
analogous to that shown in \cref{fig:SketchContour}, allowing us to
apply the general results of Supplemental
Materials~\ref{app:Integral_phase}
and~\ref{subsubsec:DerivationMonodromyGeneralContour}.
We then describe the numerical methods used to compute the QNM
spectra and test the analytic prediction.
%--------------------------------------------------------
\subsubsection{Fan--Wang black hole}\label{subsubsec:FWAnalyticCalculations}
%--------------------------------------------------------
Setting $p = 1$ in \cref{eq:metricfunctionregulargeneral} yields the Fan--Wang metric function
\begin{equation}
   B_{\rm FW}(r) =F_{\rm FW}(r)=  1-\frac{2M r^2}{(r+q)^3} = \frac{(r-r_+)(r-r_-)(r + r_3)}{(r+q)^3}\, ,
    \label{eq:FWMetricFunction}
\end{equation}
where the last equality follows from factorizing $F_{\rm FW}$ in the complex $r$-plane. 
We consider $q<q_{\rm ext}=8M/27$, for which $r_3$ is a negative real root, while $r_+$ and $r_-$ denote the outer and inner horizons, respectively.  
\begin{figure}
    \centering
    \includegraphics[width=0.55\linewidth]{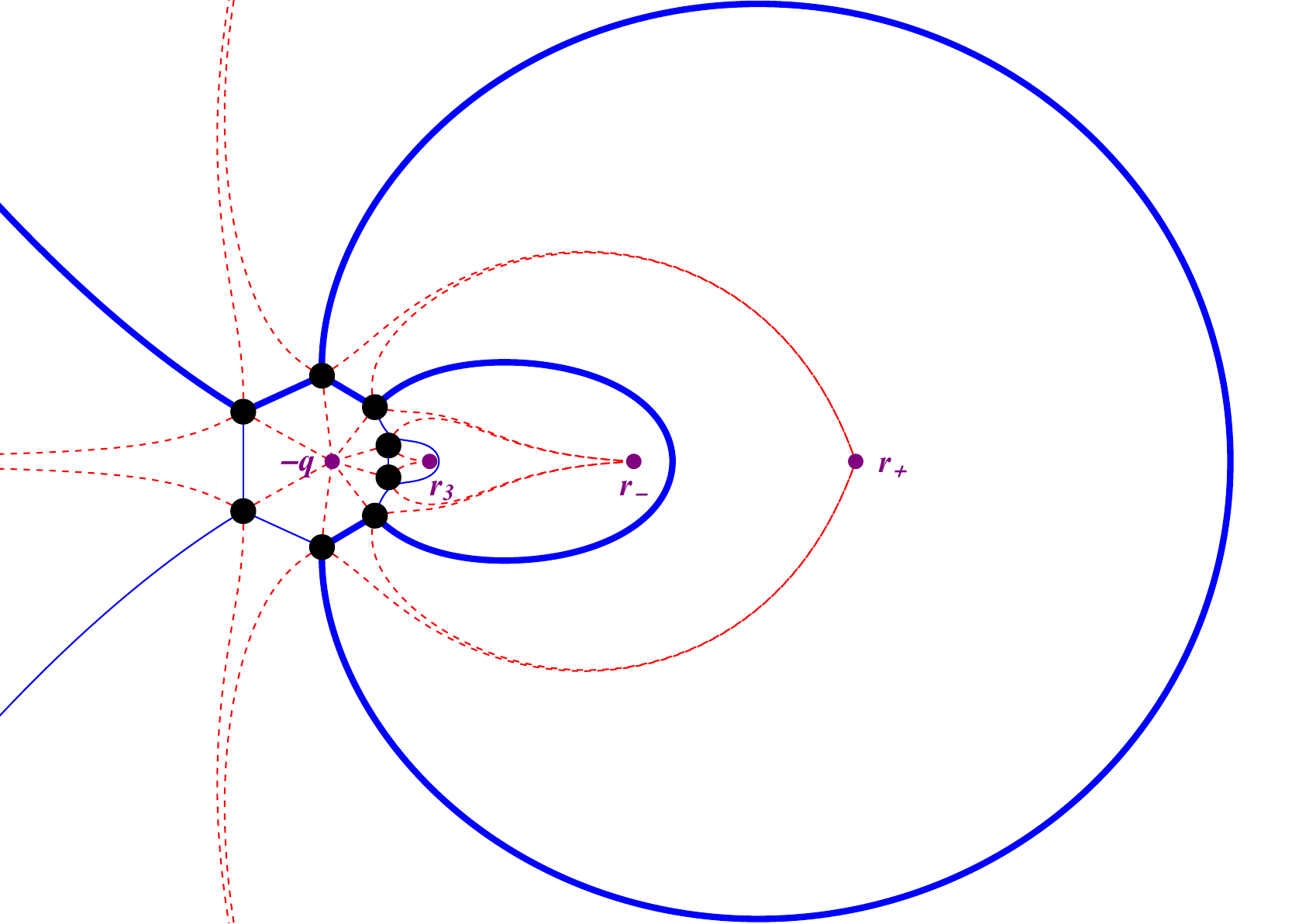}
    \caption{Network of anti-Stokes (blue) and Stokes (red dashed) lines for the Fan--Wang BH. Black dots mark the turning points of $Q^2$, while purple dots indicate the roots $r_\pm$ and $r_3$ of $F_{\rm FW}$ and its pole at $r=-q$. The thicker lines highlight the contour used for the monodromy calculation, analogous to $\Lambda$ in Fig.~\ref{fig:SketchContour}.}
    \label{fig:FanWang}
\end{figure}
The WKB function $Q^2$ follows from Supplemental Material~\ref{app:Integral_phase}. Expanding $F_{\rm FW}$ at its pole $r=-q$, we obtain
\begin{equation}
    F_{\rm FW}(r) \sim \frac{-2 M q^2}{(q+r)^3} \, .
\end{equation}
Comparing with Eq.~\eqref{eq:metric_r0}, $b(-q) = f(-q)=-2 M q^2$ and 
\begin{equation}
    N = 3 \, , \qquad \mathcal{P} = -\frac{1}{2 M q^2} \, , \qquad \alpha = \frac{15}{4}\, 
\end{equation}
from the definitions in Supplemental Material~\ref{app:Integral_phase}.
Substituting these values into \cref{eq:Q_exp_general_form} yields 
\begin{equation}
    Q_{\rm FW}^2 = \frac{\omega^2 (r+q)^6}{4 M^2 q^4} \left[1 - \frac{16 M^2 q^4}{\omega^2(r+q)^8} \right]\, .
\end{equation}
Thus, $Q_{\rm FW}^2$ has $8$ turning points clustered around $r = -q$. The resulting network of Stokes and anti-Stokes lines is shown in \cref{fig:FanWang}. 
The numerical evaluation of QNM spectra is performed in terms of $\psi(r)=(FB)^{-1/4}\, \Psi(r) = F^{-1/2} \, \Psi(r)$.
Accordingly, Eq.~\eqref{eq:EqPerturbationGeneral} becomes
\begin{equation}
    r \, F \, \psi ''+ r F' \psi' + \left(\frac{r \omega ^2}{F}-\frac{r F'+\ell(\ell+1)}{r}\right) \psi = 0\, .
    \label{eq:KGGeneral}
\end{equation}
Imposing outgoing and ingoing boundary conditions at infinity and the outer horizon~\cite{Vishveshwara:1970zz}, respectively, we require
\begin{align}
    \psi(r) &\sim r^{2M \ii \omega} \, \ee^{\ii \omega r}\, , \qquad \quad \text{for} \quad r \to \infty\, ; \label{eq:InftyBC}\\
    \psi(r)  &\sim (r-r_+)^{-\ii \sigma_+} \, , \hspace{0.61 cm}  \text{for} \quad r \to r_+\,,
    \label{eq:HorizonBC}
\end{align}
where we defined
\begin{equation}
    \sigma_+ \equiv \frac{\omega}{F_{\rm FW}'(r_+)} =  \frac{\omega  \left(r_+ +q\right){}^3}{\left(r_+-r_-\right) \left(r_+ +r_3\right)}\, .
   \label{eq:sigmaboundarydef}
\end{equation}
In the physical region $r\geq0$, the radial equation has two regular singular points at $r=r_\pm$ and an irregular singular point at infinity. 
We therefore factor out the known singular behavior of $\psi$ and expand the remaining function in a Frobenius series about the outer horizon:
\begin{equation}
    \psi(r) = \ee^{\ii \omega r} (r-r_-)^{\kappa} (r-r_+)^\lambda \sum_{j= 0}^{\infty} d_j \left(\frac{r-r_+}{r-r_-} \right)^j\,,
    \label{eq:ansatzRadial}
\end{equation}
where $\lambda = -\ii \sigma_+$ and $\kappa = 2M \ii \omega + \ii \sigma_+$. Substituting the ansatz~\eqref{eq:ansatzRadial} into the radial equation yields an $8$-term recurrence relation.

Numerical QNM frequencies are obtained with a Python code developed with the assistance of ChatGPT~6.1 and Claude~Opus~5.5. The code first reduces the recurrence relation to a three-term form through Gaussian elimination~\cite{Leaver:1990zz} and then solves it using Leaver's continued-fraction method~\cite{Leaver:1985ax}. The recurrence coefficients and continued-fraction routines were developed first, and the root-finding and mode-tracking procedure used for the highly damped spectrum was developed subsequently on top of them. We have independently validated the results up to overtone number $n=150$ against a \textsc{Mathematica} implementation of the truncated Hill-determinant method (see, e.g., Refs.~\cite{Majumdar:1989tzg,Pani:2026yzi} for details).

Highly damped modes are numerically demanding. The Gaussian elimination reducing the high-order recurrence relation to three terms suffers severe cancellations at large $|\omega|$, so we perform it in quadruple and arbitrary-precision (200-bit) arithmetic, with coefficients prepared at 60 significant digits. To ensure convergence, the truncation order of the continued fraction is up to $N_{\rm max}\sim10^5$, complemented by an asymptotic expansion of the remainder in powers of $N_{\rm max}^{-1/2}$ up to twelfth order. Overtones are obtained sequentially in $n$, using the $n$-th inversion, and labeled by increasing damping. For each mode we checked convergence against the truncation depth, the tail order, the inversion index, and the arithmetic precision. The resulting frequencies are stable to better than $10^{-10}$, except near the deepest minimum of ${\rm Re}[\omega_n]$, where they are stable to $\sim10^{-7}$. The asymptotic remainder has a branch cut along the negative imaginary $\omega$ axis, which the overtones cross near the zeros of ${\rm Re}[\omega_n]$. There, modes are followed by analytically continuing the remainder across the cut, analogously to the treatment of Schwarzschild modes near the algebraically special frequency~\cite{Leaver:1985ax,MaassenvandenBrink:2000iwh}.

%
%-----------------------------------------------------------------
\subsubsection{Bardeen black hole}
\label{subsubsec:BardeenMonodromyCalculations}
%-----------------------------------------------------------------
%
Setting $p=2$ in \cref{eq:metricfunctionregulargeneral}
yields the Bardeen metric function
\begin{equation}
    B_{\rm B}(r)=F_{\rm B}(r)
    =1-\frac{2Mr^2}{(r^2+q^2)^{3/2}}\,.
    \label{eq:Bardeen_MetricF}
\end{equation}
Unlike the Fan--Wang case, $F_{\rm B}$ is not a rational
function of $r$ and its analytic continuation introduces
branch points at $r=\pm\ii q$. Consequently, the function
$R$ in \cref{eq:R} inherits this branch structure and is
defined on a two-sheeted Riemann surface.

\begin{figure}
    \centering
    \raisebox{2.5 cm}{\includegraphics[width=0.45\linewidth]{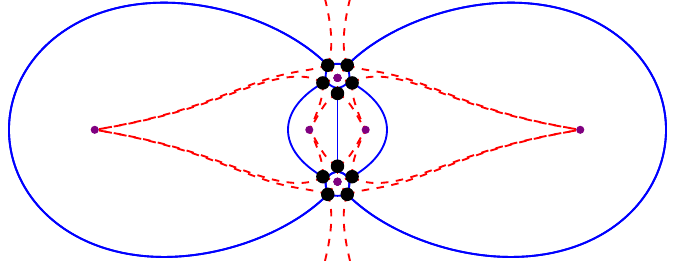}} \hfill
\includegraphics[width=0.45\linewidth]{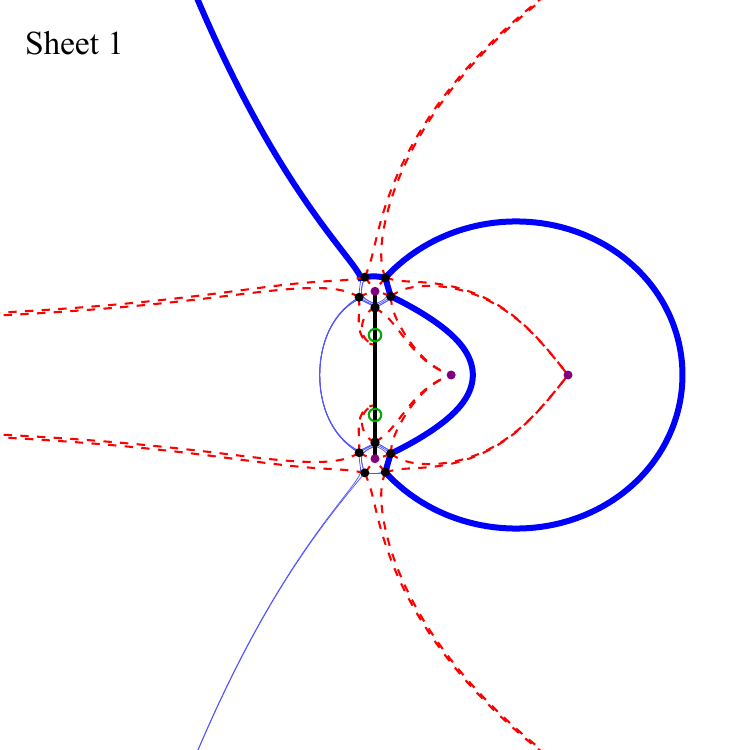} 
    \caption{Network of anti-Stokes (blue) and Stokes (dashed red) lines for the Bardeen spacetime in the complex $r$-plane. \textbf{Left panel: }network obtained without taking into account the distinction between the two Riemann sheets. \textbf{Right panel: }
    Network obtained by consistently following the analytic branch connected to the region $\text{Re}[r]>0$, with $w\sim r$ at infinity [see the discussion below \cref{eq:BardeenBranches}]. The black vertical segment joining $r=\pm\ii q$ denotes the chosen branch cut. Green circles indicate the spurious roots obtained when the zeros of $F_{\rm B}$ are mapped to the $r$-plane through $w^2 = r^2 + q^2$. The thicker blue lines highlight a monodromy contour analogous to $\Lambda$ in \cref{fig:SketchContour}. For both figures, black dots correspond to the turning points of $Q^2$. Purple dots identify the horizons and the poles of $F_{\rm B}$. } 
    \label{fig:Bardeen_r}
\end{figure}

This has important implications for the Stokes analysis:
Stokes and anti-Stokes lines must be consistently tracked
on the appropriate Riemann sheet, accounting for the
branch cuts. Neglecting this structure can lead to an
incorrect identification of the network and obscure the
connectivity discussed in Supplemental
Material~\ref{app:path_existence}.
The left panel of \cref{fig:Bardeen_r} illustrates this
issue, reproducing the anti-Stokes pattern of
Fig.~4(a) in Ref.~\cite{Lan:2022qbb}, obtained without
properly accounting for the branch structure.

A convenient analytic continuation that avoids introducing branch cuts in $F_{\rm B}$ is obtained by defining
\begin{equation}
w(r)^2 \equiv r^2 + q^2\, .
\label{eq:s_Surface}
\end{equation}
In terms of $w(r)$, the metric function \eqref{eq:Bardeen_MetricF} becomes
\begin{equation}
F_{\rm B}(r) = 1-\frac{2 M \left[w(r)^2 - q^2 \right]}{w(r)^3}\,,
\label{eq:FB_w}
\end{equation}
which is meromorphic in $w$ when analytically continued to the complex plane. The same is true for $R$, 
\begin{equation}
    R = \frac{\omega^2}{F_{\rm B}^2} - \frac{M^2 \left(w^2 - q^2\right)^2 \left(21 q^4-12 q^2 w^2-w^4\right)-M w^3 \left(w^2-q^2\right) \left[15 q^4-9 q^2 w^2+2 w^4 \ell  (\ell +1)\right]+\ell(\ell + 1)w^{10}}{w^4 \left(w^2-q^2\right) \left[2 M \left(q^2-w^2\right)+w^3\right]^2}\, .
    \label{eq:R_Bardeen_s}
\end{equation}
Following the procedure outlined in Supplemental Material~\ref{app:Integral_phase}, we construct $Q^2$ and determine its local behavior near the branch points $r=\pm \ii q$. The leading-order expansion of the metric yields 
\begin{equation}
    %f(\pm \ii q) = \ii^{\mp 3/2} M \sqrt{\frac{q}{2}} \, , \qquad 
    N = \frac{3}{2}\, , \qquad \mathcal{P} = \pm \frac{\ii^{\pm 3/2} \sqrt{2}}{M \sqrt{q}} \, , \qquad \alpha = \frac{21}{16}\, .
\end{equation}
Using \cref{eq:Q_exp_general_form}, we obtain 
\begin{equation}
    Q_{\rm B} \sim \pm \frac{\ii^{3/2} \sqrt{2}}{M \sqrt{q}} \omega \left(r \mp \ii q \right)^{3/2}\left[1 + \frac{25 M^2 q}{32 \omega^2(r\mp \ii q)^5}  \right]^{1/2}\, .
\end{equation}
Therefore, the $10$ turning points evenly distribute around $\ii q$ and $-\ii q$, as shown in \cref{fig:Bardeen_r}. 

We construct the (anti-)Stokes network using \cref{eq:R_Bardeen_s}. The relation $w^2=r^2+q^2$ defines a two-sheeted Riemann surface.
Choosing the branch cut along $r=\ii t$, with $t\in[-q,q]$,
the two branches can be written as
\begin{equation}
    w_1=r\sqrt{1+\frac{q^2}{r^2}}\,,\qquad
    w_2=-r\sqrt{1+\frac{q^2}{r^2}}\,,
    \label{eq:BardeenBranches}
\end{equation}
where the square root is defined by analytic continuation
from the positive real axis, consistently with the chosen cut.
The two sheets are distinguished by their asymptotic behavior,
$w_1\sim r$ and $w_2\sim-r$ as $r\to\infty$.
Consequently, $F_{\rm B}\sim1-2M/r$ on sheet~1, recovering
the Schwarzschild asymptotics, whereas $F_{\rm B}\sim1+2M/r$
on sheet~2.

We identify sheet~1 as the physical sheet, since it
reproduces the Schwarzschild asymptotics in the
physical region $r>0$. The corresponding (anti-)Stokes
network is shown in the right panel of \cref{fig:Bardeen_r},
with the analytic branch consistently followed throughout
the construction.
Importantly, this network admits a monodromy contour
analogous to $\Lambda$ in \cref{fig:SketchContour},
highlighted by the thicker lines. This demonstrates that
the contour required for the WKB analysis can be constructed
on the physical sheet, despite its apparent absence in the
naive single-plane representation.
The nonrational nature of the Bardeen metric function~\eqref{eq:Bardeen_MetricF} also precludes a straightforward application of Leaver's method. However, using the coordinate transformation $w \equiv \sqrt{r^2 + q^2}$, the radial wave equation~\eqref{eq:KGGeneral} becomes
\begin{equation}
    \psi \left[F_{\rm B} \left(\frac{\ell (\ell+1)}{q^2-w^2}-\frac{F_{\rm B}'}{w}\right)+\omega ^2\right] + \frac{F_{\rm B}}{w^3}\biggl\{\psi' \left[w (w^2-q^2) F_{\rm B}'+q^2 F_{\rm B}\right]+w F_{\rm B} (w^2-q^2) \psi '' \biggr\} = 0\, ,
    \label{eq:KGBardeen_scoordinate}
\end{equation}
with $F_{\rm B}$ in \cref{eq:FB_w} and with primes now denoting derivatives with respect to $w$. 

The QNM boundary conditions in the $w$ coordinate are
\begin{align}
    \psi(w) & \sim \ee^{\ii \omega w} w^{2M \ii \omega} \qquad \text{for} \quad w \to \infty\, ; \label{eq:Bardeens_BCInfty} \\
    \psi(w) &\sim (w-w_+)^{\lambda} \qquad \text{for} \quad w \to w_+\, ,
\end{align}
where we defined
\begin{equation}
    \lambda \equiv -\frac{\ii \, \omega \, w_+}{F_{\rm B}'(w_+) \sqrt{w_+^2 - q^2}}\, .
\end{equation}
We write $\psi$ factoring out the prescribed behavior at the horizon and at infinity and expanding the remaining part of the solution as a Frobenius series around the outer horizon
\begin{equation}
    \psi(w) = \ee^{\ii \omega w} \, w^\kappa (w-w_+)^\lambda \, \sum_{j= 0 }^{\infty} d_j \left(1-\frac{w_+}{w} \right)^j\, ,
\end{equation}
with the constraint $\kappa = 2M \ii \omega - \lambda$ to satisfy \cref{eq:Bardeens_BCInfty}. Plugging the ansatz into \cref{eq:KGBardeen_scoordinate} yields an $11$-term recurrence relation. 
We used the same methods as in the Fan--Wang case. The results are shown in \cref{fig:NumericalResults} for a particular value of $q$ (dot-dashed blue curve). 

To our knowledge, this is the first recurrence-based calculation
of Bardeen QNMs retaining the full metric function.
Previous Leaver-type approaches relied on small-$q$
expansions~\cite{Fernando:2012yw,Toshmatov:2015wga,Konoplya:2023ahd},
which restrict their range of validity.
Alternative methods based on WKB
approximations~\cite{Fernando:2012yw,Toshmatov:2015wga,Konoplya:2023ahd}
or direct numerical integration~\cite{Franzin:2023slm}
can face accuracy limitations or numerical instabilities
at large overtone numbers.
Our approach instead yields a finite-term recurrence relation
for the full Bardeen geometry, enabling a Leaver-type
analysis without a small-$q$ expansion.

%%-----------------------------------
\subsection{Estimate of the critical overtone number} \label{app:estimate}
%%-----------------------------------
Here we derive the estimate of the critical overtone number beyond which the spectrum tends to its asymptotic form, as discussed in the main text. We start from the general near-pole behavior in Eq.~\eqref{eq:metric_r0}. Keeping the leading coefficients, and writing $\Delta=r-r_0$, gives $H=FB\sim b(r_0)f(r_0)\Delta^{-2N}$. In the cases described by \cref{eq:Q_exp_general_form}, the leading turning-point equation is therefore
\begin{equation}
    Q^2\sim\frac{\omega^2}{b(r_0)f(r_0)}\Delta^{2N}
    -\frac{1+4\alpha}{4\Delta^2}=0\,.
    \label{eq:turning_point_balance}
\end{equation}
For $1+4\alpha\neq0$, its roots $t_k$ satisfy
\begin{equation}
    (t_k-r_0)^{2(N+1)}\sim
    \frac{(1+4\alpha)b(r_0)f(r_0)}{4\omega^2}\,.
\end{equation}
The characteristic distance of the turning points from the associated pole is $\delta r\equiv|t_k-r_0|$. This yields
\begin{equation}
    \delta r\sim
    \left[
    \frac{\sqrt{|1+4\alpha|}\,\sqrt{|b(r_0)f(r_0)|}}
    {2|\omega|}
    \right]^{1/(N+1)}\,.
    \label{eq:cluster_size_general}
\end{equation}
For $r_0\neq0$ and $z>-2$, one has $1+4\alpha=(N+1)^2$, so this becomes
\begin{equation}
    \delta r\sim
    \left[
    \frac{(N+1)\sqrt{|b(r_0)f(r_0)|}}{2|\omega|}
    \right]^{1/(N+1)}\,.
    \label{eq:cluster_size_displaced}
\end{equation}
For $r_0=0$ and $z>0$, whenever the same leading balance applies, the corresponding factor is $|N-1|$ instead of $N+1$. If $N=1$, the inverse-square coefficient in $Q^2$ vanishes and subleading terms are needed to determine the turning-point scale. The effective coefficients discussed above can likewise be inserted into \cref{eq:cluster_size_general} for the boundary cases, provided $1+4\alpha\neq0$.

We next specialize to the family in Eq.~\eqref{eq:metricfunctionregulargeneral}. On a fixed analytic sheet, let $r_0$ be a nonzero root of $r_0^p=-q^p$, with $|r_0|=q$. Expanding locally gives
\begin{equation}
    r^p+q^p=p \, r_0^{p-1}\Delta\,\left[1+{\cal O}(\Delta/r_0)\right]\,,
\end{equation}
and hence
\begin{equation}
    F=B\sim C_0\Delta^{-3/p}\,,\qquad
    C_0=-\frac{2Mr_0^2}{(p r_0^{p-1})^{3/p}}\,.
\end{equation}
In the notation of \cref{eq:metric_r0}, we have $z=N=3/p$, $b(r_0)=f(r_0)=C_0$, and
\begin{equation}
    \sqrt{|b(r_0)f(r_0)|}=|C_0|
    =\frac{2M}{p^{3/p}}q^{-1+3/p}\,.
\end{equation}
Substitution into \cref{eq:cluster_size_displaced} then gives
\begin{align}
    \delta r
    \sim\left[
    \frac{M(p+3)q^{-1+3/p}}{p^{1+3/p}|\omega|}
    \right]^{p/(p+3)}\nonumber =\frac{1}{p}\left[
    \frac{M(p+3)q^{-1+3/p}}{|\omega|}
    \right]^{p/(p+3)}\,,
    \label{eq:cluster_size_regular_family}
\end{align}
as written in the main text.

As a rule-of-thumb estimate, the transition to the asymptotic regime occurs when $\delta r\sim q$. This gives the critical frequency
\begin{equation}
    |\omega_c|\sim\frac{p+3}{p^{1+3/p}}\frac{M}{q^2}\,.
\end{equation}
Together with $|\omega_n|M\propto n$ at large overtone number, this yields \cref{eq:critical_overtone}. 

\end{document}